\documentclass[letter]{aa}

\usepackage{comment}
\usepackage{ulem}
\usepackage{color}
\usepackage{graphicx}
\usepackage{txfonts}
\usepackage{lipsum}
\usepackage[]{hyperref}
\usepackage{array}
\hypersetup{colorlinks=true,linkcolor=black,citecolor=blue}

\begin{document}

    \title{Model of X-ray and extreme-UV emission from magnetically heated atmospheres in classical T Tauri stars: Case study of TW Hya}
    
    \author{Munehito Shoda
    \inst{1},
    Riouhei Nakatani
    \inst{2}
    \and
    Shinsuke Takasao
    \inst{3}
    }
    
    \institute{
    Department of Earth and Planetary Science, School of Science, The University of Tokyo, 7-3-1 Hongo, Bunkyo-ku, Tokyo, 113-0033, Japan
    \email{shoda.m.astroph@gmail.com}
    \and
    Dipartimento di Fisica, Universit\`a degli Studi di Milano, Via Celoria, 16, I-20133 Milano, Italy
    \and
    Department of Earth and Space Science, Graduate School of Science, Osaka University, Toyonaka, Osaka 560-0043, Japan
    }
    
    \date{Received month dd, yyyy; accepted month dd, yyyy}

    \abstract
    {
    Photoevaporation caused by X-ray and UV radiation from the central star has attracted attention as a key process driving the dispersal of protoplanetary discs. Although numerous models have been used to investigate the photoevaporation process, their conclusions vary, which is partly due to differences in the adopted radiation spectra of the host star, in particular, in the extreme-UV (EUV) and soft X-ray bands. This study aims to construct the EUV and (soft) X-ray emission spectrum from pre-main-sequence stars using a physics-based model that focuses on the radiation from magnetically heated coronae. We applied a magnetohydrodynamics model capable of reproducing the coronal emission of main-sequence stars to the  pre-main-sequence star TW Hya, and we assessed its capability by comparing the predicted and observed emission line intensities. The emission lines that formed at coronal temperatures ($T = 4-13 \times 10^6$ K) are reproduced in intensity within a factor of three. Emission lines from lower-temperature ($T < 4 \times 10^6$ K) plasmas are systematically underestimated, with typical intensities at 10–30$\%$ of the observed values. This is consistent with previous findings that these emissions predominantly originate from accretion shocks. Emission lines emitted from extremely high temperatures ($T > 13 \times 10^6$ K) account for only about 1-10$\%$ of the observed values, probably because transient heating associated with flares was neglected. These results indicate that the quiescent coronal emission of pre-main-sequence stars can be adequately modelled using a physics-based approach.
    }

  \keywords{Sun: corona -- Stars: coronae --
            UV: stars --
            X-rays: stars
               }

    \titlerunning{X-ray and EUV emission from TW Hya}
    \authorrunning{M. Shoda, S. Takasao and R. Nakatani}
    
    \maketitle

\section{Introduction}

Protoplanetary discs typically persist for several million years. This is followed by a rapid dispersal in their final stages \citep{Haisch_2001_ApJ, Andrews_2020_ARAA}. In the later phase of this evolution, photoevaporation driven by high-energy radiation \citep{Hollenbach_1994_ApJ, Clarke_2001_MNRAS, Berne_2024_Sci} or its interplay with MHD winds \citep{Bai_2017_ApJ, Weder_2023_AA} was suggested to play a crucial role. Understanding disc photoevaporation is therefore crucial for determining the fate of protoplanetary discs, and consequently, planet formation \citep{Cecil_2024_AA}.

A key challenge in photoevaporation research is the lack of a consensus about the X-ray driven photoevaporation because the conclusions depend on the models. While some studies suggested that the contribution of X-ray radiation is lower than that of far-UV (FUV) and extreme-UV (EUV) radiation  \citep{Gorti_2009_ApJ_photoevaporation_rate, Wang_2017_ApJ, Nakatani_2018_ApJ_UV_and_Xray, Komaki_2021_ApJ}, others indicated that X-rays can induce significant mass loss \citep{Ercolano_2009_ApJ, Owen_2010_MNRAS, Picogna_2019_MNRAS}. The divergence in conclusions was attributed to the differences in the radiation spectra that were implemented for the model calculation and also to the treatment of thermochemistry \citep{Sellek_2022_MNRAS, Sellek_2024_AA}. Notably, the intensity of soft X-rays and the treatment of their heating efficiency were shown to significantly influence the overall mass-loss rates \citep{Gorti_2009_ApJ_time_evolution_of_PPD, Nakatani_2018_ApJ_UV_and_Xray, Sellek_2022_MNRAS, Sellek_2024_AA}. This highlights the need to accurately input radiation spectra in the soft X-rays and adjacent EUV bands. However, these radiations are strongly absorbed by interstellar and circumstellar materials, which makes observational spectral acquisition challenging and contributes to the uncertainties.

\begin{figure}[!t]
\centering
\includegraphics[width=75mm]{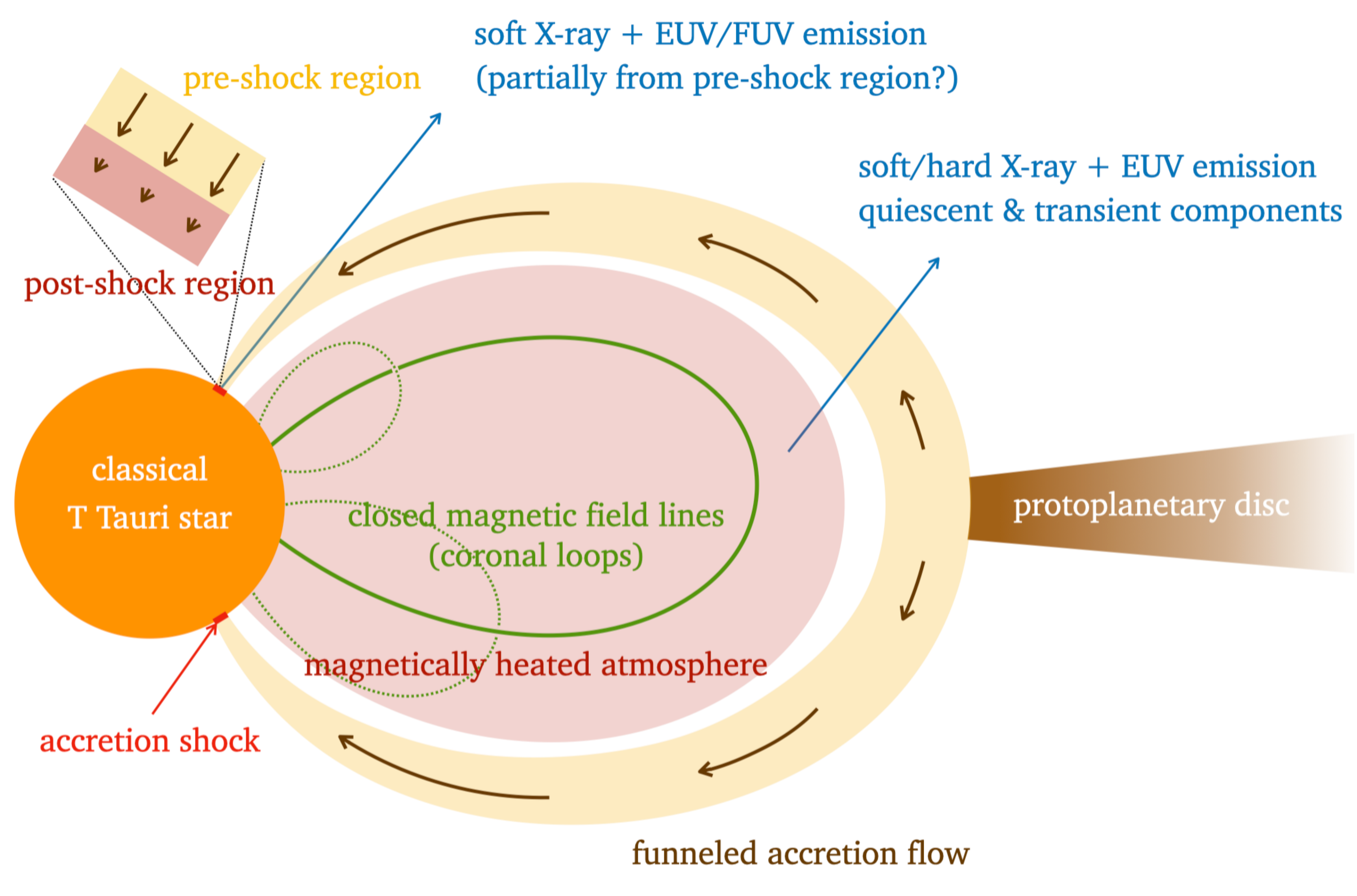}
\caption{
Conceptual picture of X-ray, EUV, and FUV emissions from classical T Tauri stars.}
\label{fig:schematic_picture}
\end{figure}

X-ray and EUV emissions from classical T Tauri stars (CTTSs) originate from two sources: the downstream of accretion shock, and the magnetically heated atmosphere  (transition region and corona) \citep[Fig. \ref{fig:schematic_picture}, see also][]{Argiroffi_2007_AA, Gunther_2007_AA}. The atmospheric emission is further classified into the quasi-steady emission from a quiescent corona and the transient emission by flaring activities \citep{Gudel_2004_AARv}. X-rays in the majority of CTTSs predominantly arise from the corona \citep{Robrade_2006_AA, Stassun_2007_ApJ}, and accretion shocks only contribute to the excess in the soft component \citep{Lamzin_1999_AstL, Telleschi_2007_AA_Xray_Tau_AU_complex}, partially because of the effective absorption of the X-ray by the pre-shock accretion streams \citep{Lamzin_1996_AA, Colombo_2019_AA_preshock_heating}. Although they were not directly observed, accretion shocks are anticipated to contribute to the EUV emission as well, considering that accretion shocks can dominate the FUV emission \citep{Johns-Krull_2000_ApJ, Hinton_2022_ApJ}. Thus, appropriately modelling the emission from the magnetically heated atmosphere and the (downstream of) accretion shock is essential for the physics-based construction of a CTTS emission spectrum.

A number of physics-based models have been proposed to reproduce the high-energy emission from accretion shocks \citep{Sacco_2008_AA, Colombo_2016_AA, de_Sa_2019_AA}. Although several coronal models for CTTSs have been proposed \citep{Cranmer_2009_ApJ, Cohen_2023_ApJ}, however, no models have quantitatively reproduced the observed coronal (X-ray) emission lines from a CTTS by directly comparing individual emission lines between observations and models. Due to the absence of quantitatively tested model spectra, the empirical emission spectra derived from X-ray observations were used \citep{Ercolano_2009_ApJ, Ercolano_2021_MNRAS} to discuss the contributions of soft X-rays and EUV.

We model the coronal emission of CTTSs using a physics-based model. We use a time-dependent magnetohydrodynamics model capable of reproducing the quiescent X-ray and EUV emissions of solar-type stars \citep{Shoda_2021_AA, Shoda_2024_AA}. Several studies have quantitatively discussed the reproducibility of the emission lines of the solar corona with a similar model \citep{Oran_2013_ApJ, Sachdeva_2021_ApJ, Shi_2024_ApJ}. However, the model presented here adopts the photosphere as the boundary without employing a mean-field description of Alfv\'en-wave heating, in contrast to available three-dimensional global coronal models \citep{van_der_Holst_2014_ApJ, Downs_2021_ApJ, Parenti_2022_ApJ}.
Specifically, we apply the model to TW Hya. Although the X-ray emissions from TW Hya predominantly originate from the low-temperature plasma ($\log T \le 6.5$) in the downstream region of accretion shocks \citep{Kastner_2002_ApJ, Stelzer_2004_AA, Argiroffi_2017_AA}, several emission lines from high-temperature plasma, which cannot be reproduced by accretion shocks alone, have been observed \citep{Brickhouse_2010_ApJ}. These observations facilitate the evaluation of our model.

\section{Model description \label{sec:model_description}} 

We modelled a coronal loop as a one-dimensional MHD system. An overview of the model is provided in Appendix~\ref{app:model_overview}. The simulation requires specifying the surface physical parameters as inputs, which are derived from the stellar observable parameters. The mass ($M_\ast$) and radius ($R_\ast$) of TW Hya are estimated to be $M_\ast/M_\odot = 0.5-0.8$ and $R_\ast/R_\odot = 0.8-1.1$, respectively \citep{Debes_2013_ApJ, Argiroffi_2017_AA}. However, near-infrared observations reveal $\log g = 4.2$, with $g$ denoting the surface gravity in cgs units \citep{Sokal_2018_ApJ}. To align with this result, $M_\ast$ and $R_\ast$ were set accordingly as 
\begin{align}
    M_\ast/M_\odot = 0.7, \hspace{1em} R_\ast/R_\odot = 1.1.
\end{align}
The effective temperature ($T_{\rm eff}$) and mean surface field strength ($\langle B \rangle_\ast$) were derived from high-resolution near-infrared observations as follows \citep{Sokal_2018_ApJ}: 
\begin{align} 
    T_{\rm eff} = 3.8 \times 10^3 {\rm \ K}, \hspace{1em} \langle B \rangle_\ast = 3.0 \times 10^3 {\rm \ G}.
\end{align} 
Using the observed values ($M_\ast$, $R_\ast$, $\langle B \rangle_\ast$, $T_{\rm eff}$) listed above, we determined the surface parameters as detailed in Appendix~\ref{app:surface_parameters}.

In addition to the surface parameters, the atmospheric elemental abundance is required to calculate the radiative cooling function and the emission spectrum. While X-ray observations provide constraints on elemental abundances, significant uncertainties persist. The abundance of Ne in the TW Hya atmosphere varies extensively, from one to ten times the solar value \citep{Stelzer_2004_AA, Raassen_2009_AA, Brickhouse_2010_ApJ}. This variation may result from technical factors, such as inversion methods and observational instruments, and from intrinsic uncertainties associated with the stellar activity \citep{Gudel_2004_AARv, Testa_2010_SSRv}. Considering the difficulty in determining the most accurate estimates, we adopted the median value reported in Model C from \citet{Brickhouse_2010_ApJ}. For unspecified elements, the solar photospheric values \citep{Anders_1989_GeCoA} were used. 

We fixed the half-loop length ($L_{\rm half}$) at $L_{\rm half} = 80 {\rm \ Mm}$ (see Fig.~\ref{fig:model_schematic} for the definition). This value was chosen because $L_{\rm half} = 80 {\rm \ Mm}$ best explains solar observations when the loop length is fixed to a specific value \citep{Shoda_2024_AA}. Since the model requires a long time to reach temporal convergence, we first performed a low-resolution calculation (minimum grid size: $\Delta s_{\rm min} = 20 {\rm \ km}$) for 5000 minutes, followed by a high-resolution calculation ($\Delta s_{\rm min} = 1 {\rm \ km}$) for 1500 minutes. The time step was typically $\Delta t = 3-4 \times 10^{-5} {\rm \ s}$, and the high-resolution calculation required 2.7 billion time steps. The time-averaged data were computed from the last 500 minutes, a duration far greater than the maximum wave period applied at the boundary (76 minutes). Corrections associated with the use of the LTRAC method were considered to calculate the (line-of-sight) differential emission measure \citep[see Eq. (23) in][]{Iijima_2021_ApJ}.

\begin{figure}[!t]
\centering
\includegraphics[width=65mm]{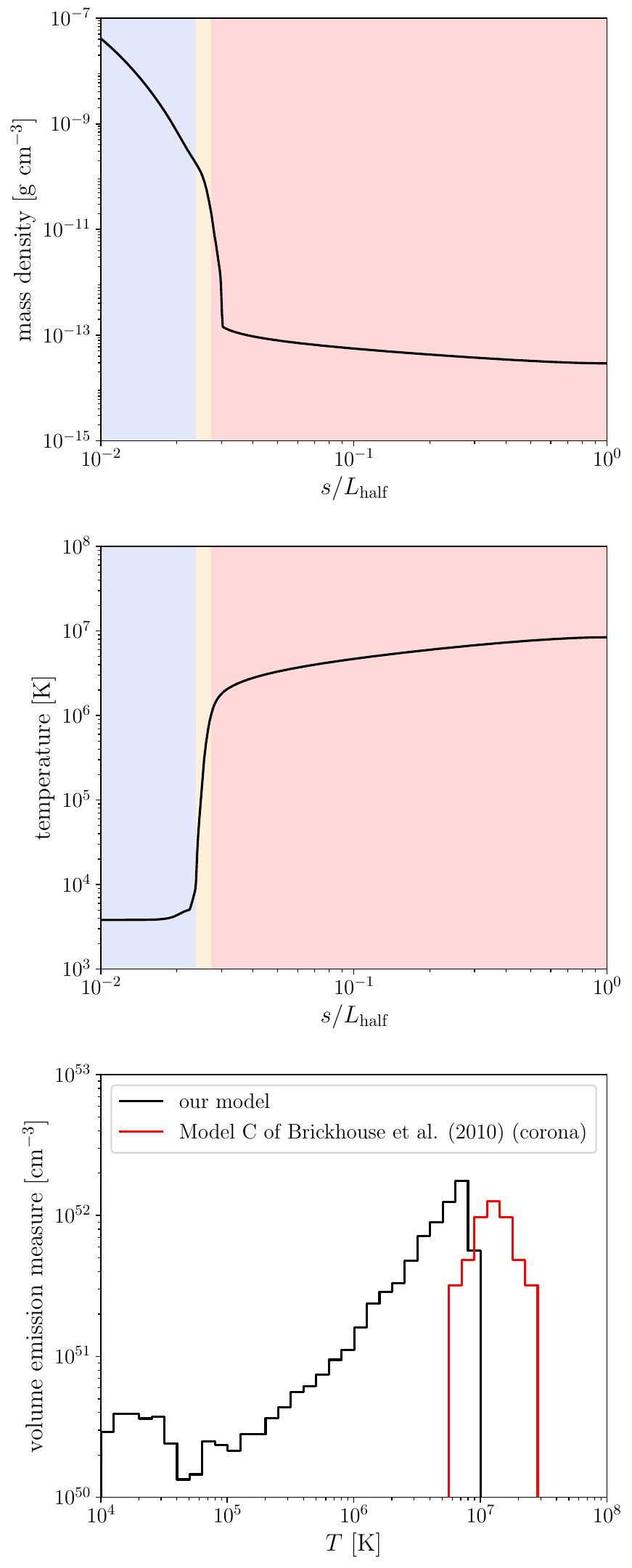}
\caption{
Time-averaged simulation results. The top panel illustrates the density along the loop, and the middle panel shows the temperature distribution along the loop. The shaded blue, orange, and red areas represent the chromosphere ($T < 10^4 {\rm \ K}$), transition region ($T = 10^{4-6} {\rm \ K}$), and corona ($T > 10^{6} {\rm \ K}$), respectively. The bottom panel presents the volume emission measure distribution, and the red line represents the coronal emission measure inferred from X-ray observations \citep[Model C in][]{Brickhouse_2010_ApJ}.
}
\label{fig:simulation_result_overview_binned_volume_emd}
\end{figure}

\begin{figure}[!t]
\centering
\includegraphics[width=70mm]{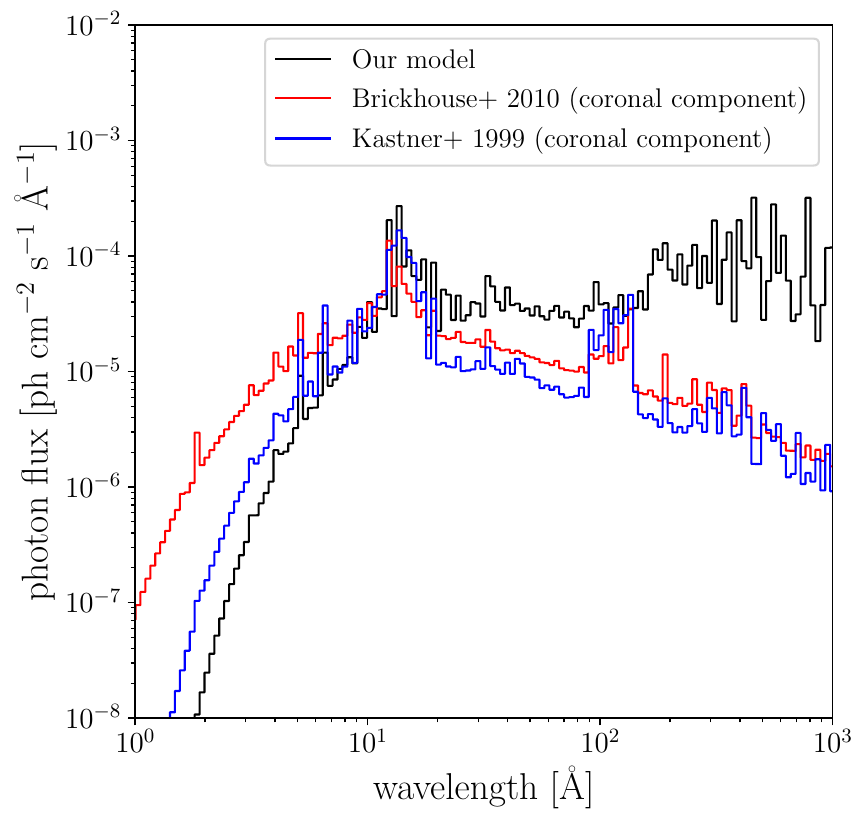}
\caption{
Comparison of the modelled photon-flux spectrum (black) with the spectra derived from the coronal emission measures obtained by X-ray observations (blue: \citet{Kastner_1999_ApJ}, and red: \citet{Brickhouse_2010_ApJ}). 
}
\label{fig:spectrum_overview}
\end{figure}

\section{Result \label{sec:simulation_result}}

Figure~\ref{fig:simulation_result_overview_binned_volume_emd} presents the time-averaged simulation results. The figure displays from top to bottom the density distribution along the loop, the temperature distribution along the loop, and the volume emission measure as a function of temperature. The volume emission measure (${\rm EM}_{\rm V}$) is related to the line-of-sight emission measure (${\rm EM}_{\rm los}$), directly computed from the simulation, by the following relation \citep{Dennis_2024_SoPh}: 
\begin{align}
    \frac{F_\lambda}{{\rm EM}_{\rm V}} = \frac{I_\lambda}{{\rm EM}_{\rm los}} \times \frac{1}{d^2},
\end{align}
where $F_\lambda$ and $I_\lambda$ represent the flux and intensity at wavelength $\lambda$, respectively, and $d = 59.5\ {\rm pc}$ is the distance to TW Hya \citep{Gaia_2016_AA}. Assuming that the corona is entirely filled with coronal loops, compatible with the low filling factor of accretion streams in TW Hya \citep[$\sim 0.3 \%$,][]{Ingleby_2013_ApJ}, the flux and intensity are related by $F_\lambda = \pi I_\lambda \left( R_\ast/d \right)^2$ \citep{Shoda_2024_AA}. Subsequently, ${\rm EM}_{\rm V}$ can be obtained from ${\rm EM}_{\rm los}$ as
\begin{align}
    {\rm EM}_{\rm V} = \pi R_\ast^2 {\rm EM}_{\rm los}.
\end{align}
The spectrum was computed using an emission measure with a temperature resolution of 0.01 dex. However, in comparing with the emission measure estimated in the literature \citep[Model C of ][]{Brickhouse_2010_ApJ}, we present the emission measure degraded to a resolution (bin size) of 0.1 dex. We note that the value of ${\rm EM}_{\rm V}$ changes in proportion to the bin size.

\begin{figure*}[!t]
\centering
\includegraphics[width=160mm]{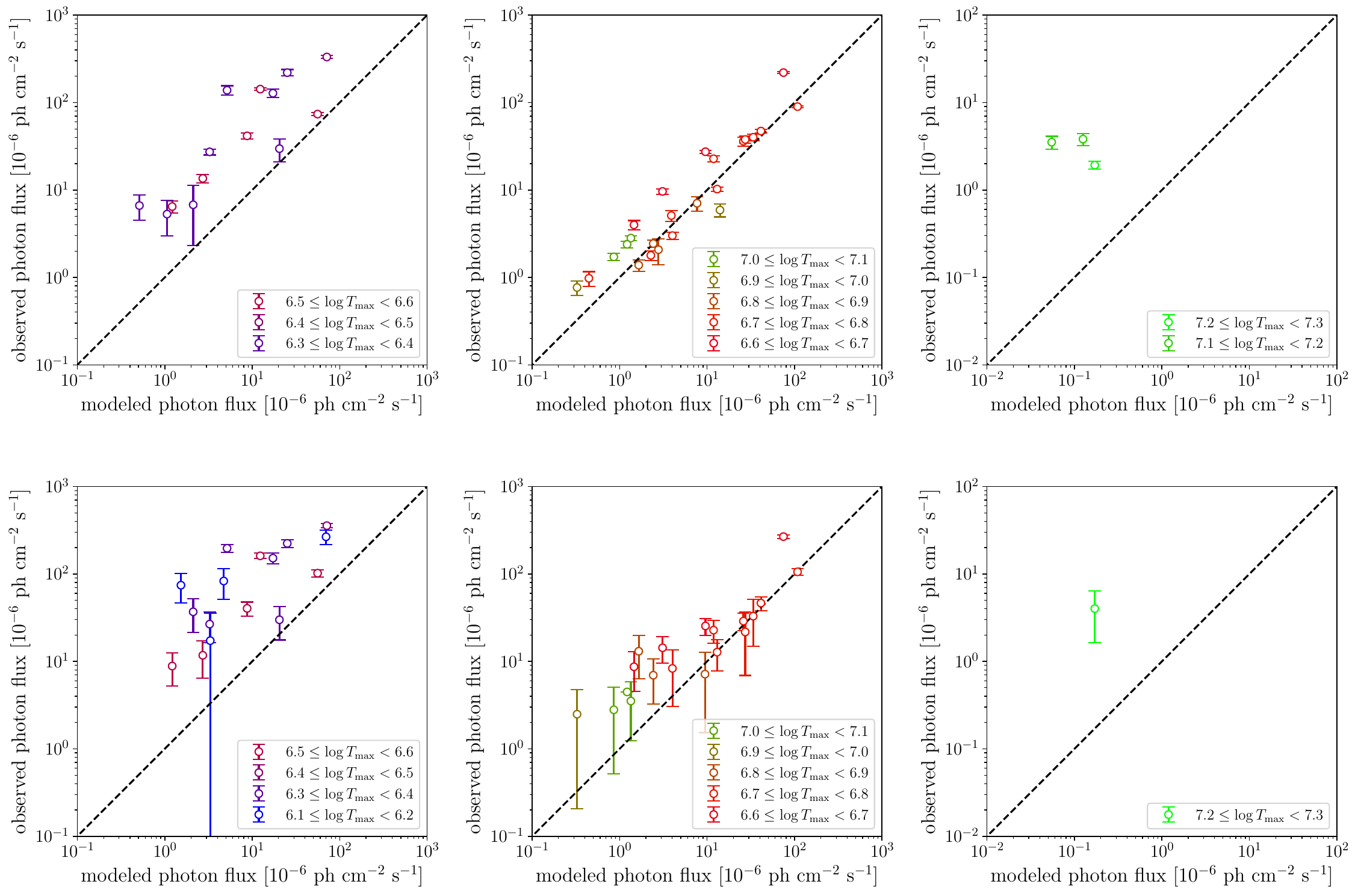}
\caption{
Comparison of the modelled emission-line intensities with the absorption-corrected observed values. The top panels compare the model results with those from \citet{Brickhouse_2010_ApJ}, and the bottom panels compare them with \citet{Raassen_2009_AA}. The panels from left to right correspond to emission lines formed at low ($\log T_{\rm max} <6.6$), medium ($6.6 < \log T_{\rm max} < 7.1$), and high temperatures ($\log T_{\rm max} > 7.1$), respectively. The error bars represent the 1$\sigma$ statistical errors listed in Table~\ref{table:emission_line_list}.
}
\label{fig:emission_lines_comparison_with_absorption_correction}
\end{figure*}

As shown by the density and temperature distributions, the modelled atmosphere of TW Hya consists of a cool chromosphere and a hot corona that are separated by a thin transition region. The maximum coronal temperature at each time ranges from $\log T = 6.87-6.99$ ($k_{\rm B} T = 0.637-0.840 {\rm \ keV}$), with an average of $\log T = 6.93$ ($k_{\rm B} T = 0.727 {\rm \ keV}$). The high-temperature (coronal) component of the two-temperature emission measure is observed within the range of $k_{\rm B} T = 0.76-1.00\ {\rm keV}$ \citep{Kastner_1999_ApJ}, which is slightly higher than our result. This difference probably results from the omission of flares in our model, where high-temperature ($10^{7-9}$ K) plasma is generated \citep{Getman_2008_ApJ_I, Getman_2008_ApJ_II, Getman_2021_ApJ}. 

Figure~\ref{fig:spectrum_overview} presents the model spectrum calculated from the time-averaged emission measure with the CHIANTI atomic database version 10 \citep{Dere_1997_AA, Del_Zanna_2021_ApJ}. The solid black line illustrates the modelled photon-flux spectrum as a function of wavelength. The red and blue lines represent the spectra derived from the coronal (high-temperature) emission measure inferred from X-ray observations, where the red lines correspond to Model C from \citet{Brickhouse_2010_ApJ} and the blue lines to the ROSAT data by \citet{Kastner_1999_ApJ}. These spectra were also calculated using CHIANTI version 10. 

Although the three spectra agree in the observable X-ray range (10-20 ${\rm \AA}$, 0.6–1.2 keV), at longer wavelengths, in particular, in the EUV range ($ > 100 {\rm \ \AA}$), the model spectrum is significantly higher than those derived from the observational emission measures. For the photon emission rate in the EUV range ($\Phi_{\rm EUV}$), our model predicts $\Phi_{\rm EUV} = 3.3 \times 10^{40} {\rm \ s}^{-1}$, whereas the observational spectra indicate values 20 times lower ($\Phi_{\rm EUV} \sim 1.5 \times 10^{39} {\rm \ s}^{-1}$). Nonetheless, this does not imply an inconsistency between the model and observations. The spectra derived from observations were calculated using the emission measure obtained from X-ray data, which neglects the cooler component of the emission measure that significantly impacts the EUV emission. Therefore, the observed discrepancy suggests that inferring the EUV spectra from X-ray observations is a limited exercise.

We show in Fig.~\ref{fig:emission_lines_comparison_with_absorption_correction} the scatter plots comparing observed and modelled photon fluxes across three temperature ranges: $\log T_{\rm max}<6.6$, $6.6<\log T_{\rm max} < 7.1$, and $\log T_{\rm max}>7.1$. Since the fluxes reported in the literature were not corrected for absorption, we applied a correction using the absorption model by \citet{Wilms_2000_ApJ}, assuming a hydrogen column density of $N_{\rm H} = 1.0 \times 10^{21} {\rm \ cm^{-2}}$ \citep{Brickhouse_2010_ApJ}. The upper three panels of Fig.~\ref{fig:emission_lines_comparison_with_absorption_correction} compare our model with that of \citet{Brickhouse_2010_ApJ}, and the lower three panels compare our model with \citet{Raassen_2009_AA}.  The error bars represent the $1\sigma$ statistical errors reported in each study. 

Comparison of fluxes for emission lines formed at low temperatures ($\log T_{\rm max} < 6.6$) indicates that the model values are systematically lower than observed. This discrepancy aligns with the view that the low-temperature emission lines are primarily generated by the accretion shocks \citep{Kastner_2002_ApJ, Argiroffi_2017_AA}. The maximum temperature of the accretion shock ($T_{\rm acc}$) is estimated as \citep[see e.g. ][]{Schneider_2022}
\begin{align} 
    T_{\rm acc} = 5.4 \times 10^{6} {\rm \ K} \left( \frac{M_\ast}{M_\odot} \right) \left( \frac{R_\ast}{R_\odot} \right)^{-1}. 
\end{align}
Using the values for TW Hya ($M_\ast/M_\odot = 0.7$, $R_\ast/R_\odot = 1.1$), we find $\log T_{\rm acc} = 6.54$, which is close to the low-temperature line threshold ($\log T_{\rm max} = 6.6$) in this analysis. 

In the moderate temperature range ($6.6 < \log T_{\rm max} < 7.1$), the emission-line intensities from our model and observations align closely. This is consistent with the conventional understanding that the emission in this temperature range originates from the corona rather than the accretion shock. Notably, our model did not retrospectively calculate the emission measure to fit these emission lines; instead, it used prescribed magnetic fluxes and abundances to model the emission lines in a forward manner. Replication of the emission lines in this temperature range indicates that our model accurately describes the corona and its emissions in TW Hya.

The emission lines from high-temperature plasma ($7.1 < \log T_{\rm max}$) are underestimated by one to two orders of magnitude. This trend is also observed in modelling highly active solar-type stars \citep{Shoda_2024_AA} and is probably due to the omission of transient large-scale heating events that produce flares. Observations suggest that flares can produce plasma at temperatures of $10^{7-9}$ K, implying that models incorporating flares \citep{Waterfall_2019_MNRAS, Kimura_2023_ApJ} could better replicate emission lines from these high-temperature plasmas. 

The discussion so far relies on the data positions in the scatter plot, making it relatively qualitative. Nevertheless, a quantitative assessment of model accuracy at each temperature range yields the same conclusion (see Appendix~\ref{app:model_accuracy_vs_temperature} for further details).

\section{Summary and discussion}

We applied a model that reproduced the X-ray and EUV emission of solar-type stars to a pre-main-sequence star (TW Hya) to examine the feasibility of reproducing the coronal emission of pre-main-sequence stars. By comparing the photon flux of emission lines between the model and observations, we confirmed that the model successfully reproduced the observed emission lines originating from the corona, with a formation temperature of $\log T_{\rm max} = 6.6-7.1$. This suggests that the same coronal heating mechanism in which magnetic fields are stirred by surface convection to heat the atmosphere is valid for both pre-main-sequence and main-sequence stars, indicating the potential for physics-based modelling of coronal emissions based on this framework. Our model fails to explain radiation from high-temperature plasma ($\log T_{\rm max} > 7.1$), however, and a more comprehensive model including flares is required for a proper evaluation of its validity.

The coronal emission predicted in this study is higher by one to two orders of magnitude in EUV intensity than estimates from X-ray observations. This indicates that the role of EUV in photoevaporation may have been undervalued, suggesting the need to reassess EUV-driven photoevaporation using realistic spectra \citep{Nakatani_2024_ApJ}. The estimated EUV emission rate ($3\times10^{40}\,\mathrm{s^{-1}}$) is high enough to potentially yield mass-loss rates of $\sim 10^{-9}\,M_\odot \,\mathrm{yr^{-1}}$ \citep{Tanaka_2013_ApJ}, which are comparable to those caused by FUV and X-ray photoevaporation \citep{Komaki_2021_ApJ, Sellek_2024_AA}. The EUV emission rate in our model, however, which serves as a lower limit due to the exclusion of the accretion shock, exceeds the observed upper limit of the EUV emission rate reaching the disc \citep[$\sim 1.5 \times10^{40}\,\mathrm{s^{-1}}$,][]{Pascucci_2014_ApJ}. The discrepancy may be attributed to the optical thickness of the free-free emission from TW Hya (at least in part), which should be verified in future research.

Care must be taken when interpreting the findings of this study. The model we employed is one-dimensional and includes several unavoidable free parameters. The most critical of these is the coronal loop length \citep{Takasao_2020_ApJ}, as prior research has shown that coronal radiation can vary by several factors depending on the loop length \citep{Shoda_2024_AA}. The turbulence that causes coronal heating was modelled phenomenologically \citep{Cranmer_2009_ApJ, Shoda_2018_ApJ_a_self-consistent_model}. While this model is reasonably valid for the solar corona, caution must be exercised when applying it to pre-main-sequence stars with much stronger magnetic fields than the Sun. In addition, the applicability of this model should be discussed considering the characteristics of accretion in CTTSs. In the case of TW Hya, we have argued that the emissions of low formation temperature is dominated by accretion shock, given that TW Hya exhibits a strong continuous accretion \citep[for example,][]{Herczeg_2023_ApJ}. However, if accretion varies significantly over time due to stellar winds \citep{Cohen_2023_ApJ}, the resulting emission may weaken, enabling this model to reproduce the total emission spectrum of CTTSs.

Despite the aforementioned caveats, it is clear that the physical-based estimation of X-ray and EUV spectrum has reached a sufficiently feasible stage even for pre-main-sequence stars. Our physical model represents a significant step towards a more accurate understanding of photoevaporation and more reliable spectral reconstruction from observational data.

\section*{Data availability}
The numerical data of the emission spectrum and differential emission measure can be accessed via GitHub: \url{https://github.com/munehitoshoda/coronal_spectrum_TWHya}. The emission line data are also summarized in Appendix~\ref{app:emission_line_list}.

\begin{acknowledgements}
Numerical computations were carried out on the Cray XC50 and Wisteria/BDEC-01 Odyssey at the University of Tokyo. This work is supported by JSPS KAKENHI Grant Numbers JP24K00688 (MS), JP22K14074, JP21H04487, JP22KK0043 (ST). RN acknowledges support from the European Union (ERC Starting Grant DiscEvol, project number 101039651) and from Fondazione Cariplo, grant No. 2022-1217. Views and opinions expressed are, however, those of the author(s) only and do not necessarily reflect those of the European Union or the European Research Council. Neither the European Union nor the granting authority can be held responsible for them. This work is supported by “Joint Usage/Research Center for Interdisciplinary Large-scale Information Infrastructures (JHPCN)” in Japan (Project ID: jh230046). This work made use of matplotlib, a Python library for publication quality graphics \citep{Hunter_2007_CSE}, and NumPy \citep{van_der_Walt_2011_CSE}.
\end{acknowledgements}

\bibliographystyle{aa}

\appendix

\section{Model overview \label{app:model_overview}}

\begin{figure}[!h]
\centering
\includegraphics[width=90mm]{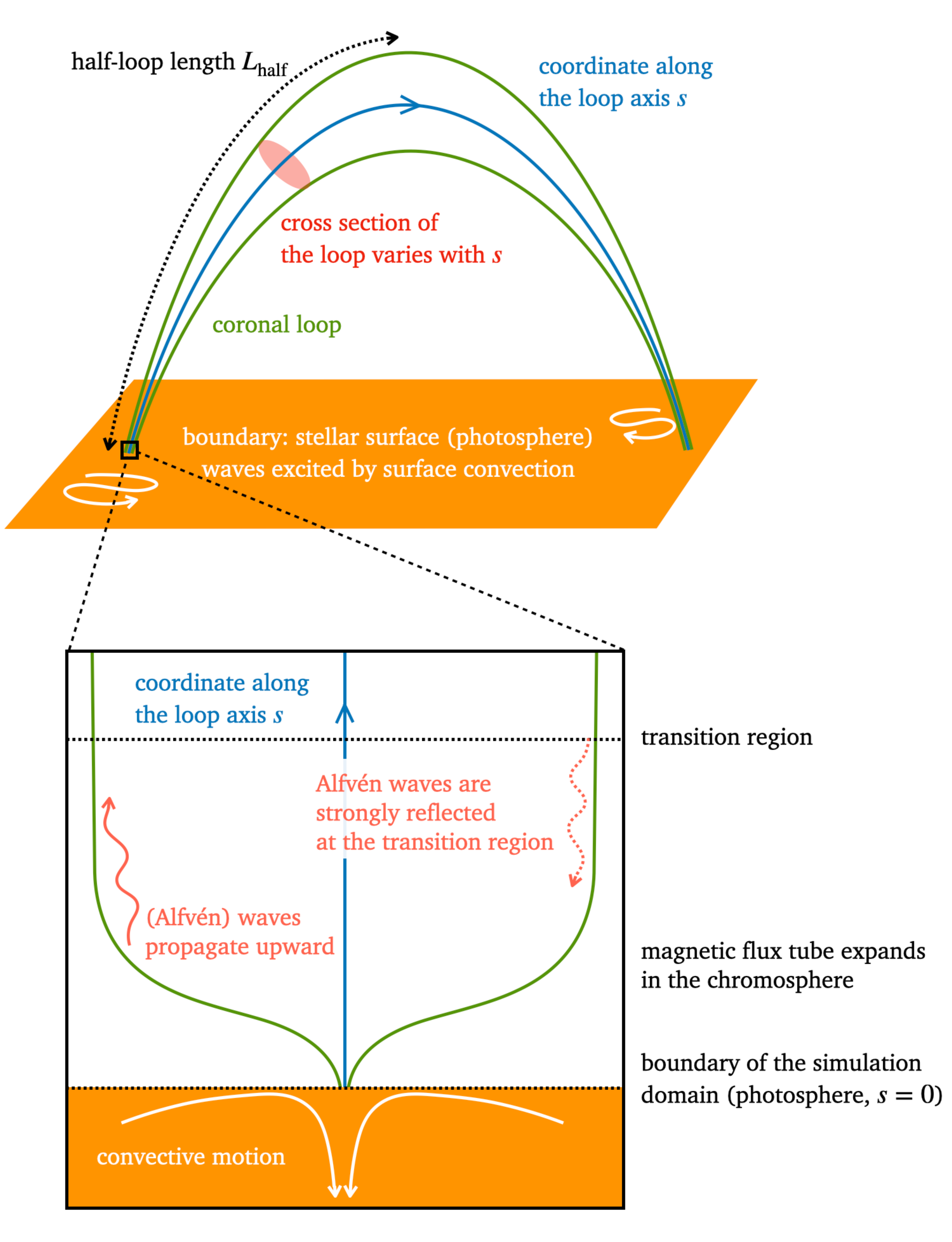}
\caption{
A schematic representation of our model.
}
\label{fig:model_schematic}
\end{figure}

In this work, a coronal loop is modelled as a one-dimensional tube, a method frequently employed, particularly in the classical studies of coronal loops and associated physical processes \citep{Peres_1982_ApJ, Hansteen_1993_ApJ, Ofman_2002_ApJ, Bradshaw_2006_AA}. Figure~\ref{fig:model_schematic} illustrates a schematic representation of our model. One unique feature of our model, as highlighted in the bottom part of Fig.~\ref{fig:model_schematic}, is that it considers the expansion of the magnetic field within the chromosphere. To model energy injection into the atmosphere, we impose velocity and magnetic field perturbations at the boundary that mimics the convective motion in the photosphere.

The basic equations employed are the same as those in \citet{Shoda_2024_AA}, with the exception of using the transition-region broadening technique \citep[LTRAC method,][]{Iijima_2021_ApJ}. We solve the one-dimensional MHD equations in an expanding flux tube, incorporating gravity, thermal conduction, and radiative cooling. The turbulent dissipation is modelled phenomenologically to address coronal heating in the one-dimensional framework. For further details, refer to \citet{Shoda_2021_AA} and \cite{Shoda_2024_AA}.

\section{Estimation of surface parameters \label{app:surface_parameters}}

\begin{figure}[!t]
\centering
\includegraphics[width=80mm]{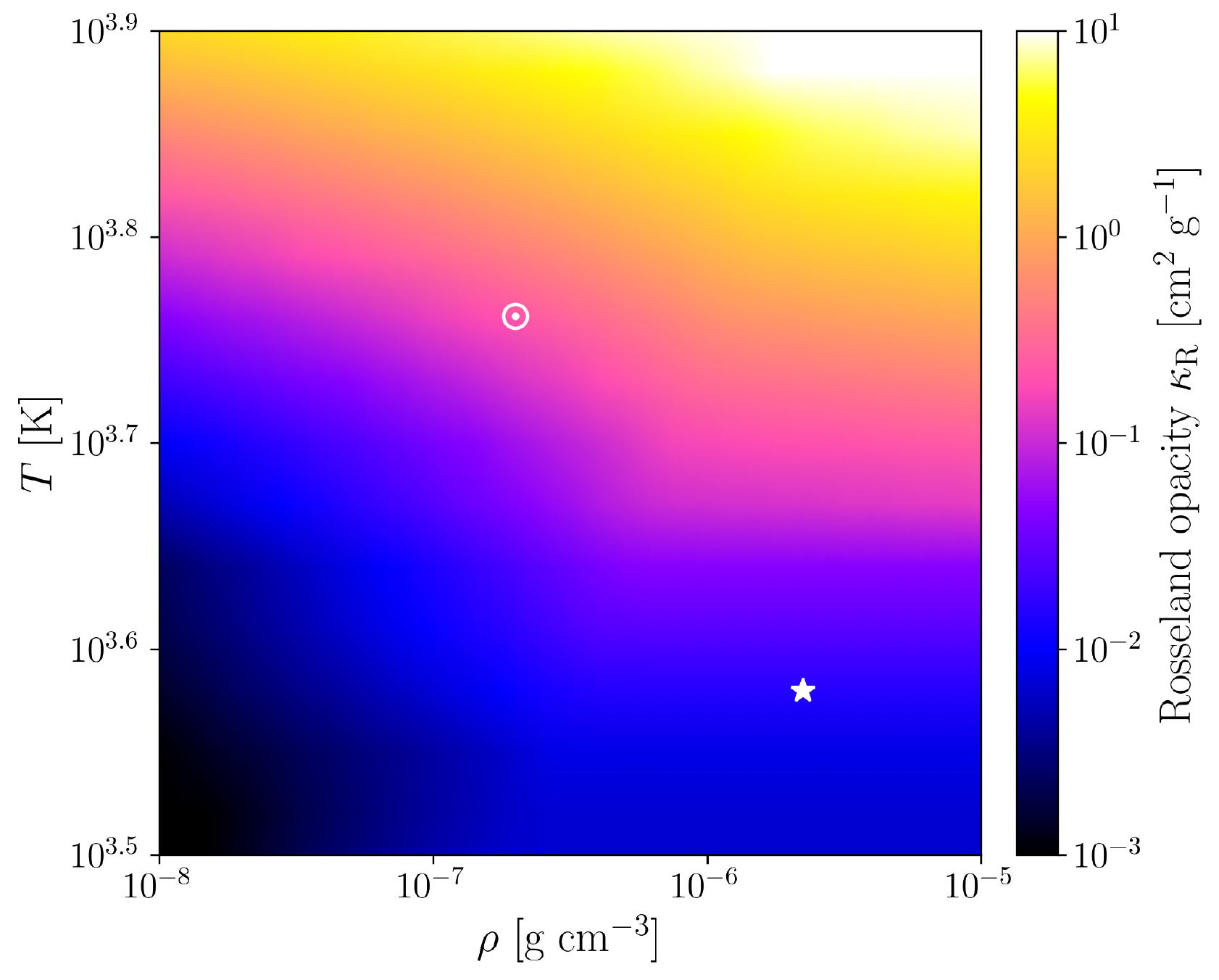}
\caption{
Rosseland opacity table used in this study, represented as a function of mass density and temperature. The symbol $\odot$ denotes the estimated values for the solar photosphere, while the asterisk corresponds to the estimated values for the photosphere of TW Hya.
}
\label{fig:Rosseland_opacity}
\end{figure}

In our model, the stellar surface is defined as the boundary, necessitating the prescription of stellar surface physical quantities from observational parameters. Defining the photosphere at the Rosseland optical depth of 2/3 \citep{Kippenhahn_1990_book}, the photospheric density ($\rho_\ast$) satisfies the following relation:
\begin{align}
    \rho_\ast H_\ast \kappa_{\rm R} (\rho_\ast, T_\ast) = \frac{2}{3},
\end{align}
where $H_\ast$ denotes the pressure scale height in the photosphere. $\kappa_{\rm R}$ represents the Rosseland opacity, and was calculated using three tables from the literature \citep{Ferguson_2005_ApJ, Seaton_2005_MNRAS}, connected by bilinear interpolation (see Fig.~\ref{fig:Rosseland_opacity})

Determining the photospheric density allows for the estimation of stellar surface convection parameters. The typical convective velocity ($\delta v_\ast$) is estimated assuming that the energy transport rate in the convection zone equals the stellar luminosity \citep{Suzuki_2018_PASJ, Sakaue_2021_ApJ}. Consequently, $\delta v_\ast$ is determined to satisfy the following relationship:
\begin{align}
    \rho_\ast \delta v_\ast^3 \propto \frac{L_\ast}{4 \pi R_\ast^2} = \sigma_{\rm SB} T_\ast^4, \label{eq:deltav_assumption}
\end{align}
where $L_\ast$ is the stellar luminosity and $\sigma_{\rm SB}$ is the Stefan-Boltzmann constant. It is crucial to note that the equality $\rho_\ast \delta v_\ast^3 = \sigma_{\rm SB} T_\ast^4$ does not hold since the kinetic energy flux is no longer dominant at the stellar surface among all energy flux terms \citep{Nordlund_2009_LRSP}. From Equation~\eqref{eq:deltav_assumption}, $\delta v_\ast$ can be derived as follows:
\begin{align}
    \delta v_\ast = \delta v_\odot \left( \frac{\rho_\ast}{\rho_\odot} \right)^{-1/3} \left( \frac{T_\ast}{T_\odot} \right)^{4/3},
\end{align}
where the subscript $\odot$ indicates the solar values.
Specifically, we set $\rho_\odot = 1.83 \times 10^{-7} {\rm \ g \ cm^{-3}}$ and $\delta v_\odot = 1.06 \times 10^5 {\rm \ cm \ s^{-1}}$.

The horizontal correlation length of surface convection, $\lambda_\perp$, is assumed to scale with the pressure scale height, such that 
\begin{align}
    \lambda_{\perp, \ast} = \lambda_{\perp, \odot} \frac{H_\ast}{H_\odot},
\end{align}
where $\lambda_{\perp, \odot} = 150 {\rm \ km}$. The typical frequency of surface convection, $f^{\rm conv}_\ast$, is assumed to be inversely proportional to the convection turnover time ($\lambda_{\perp, \ast}/\delta v_\ast$), resulting in
\begin{align}
    f^{\rm conv}_\ast = f^{\rm conv}_\odot \left( \frac{\lambda_{\perp,\ast}}{\lambda_{\perp,\odot}} \right)^{-1} \left( \frac{\delta v_\ast}{\delta v_\odot} \right),
\end{align}
where $f^{\rm conv}_\odot = 1.0 \times 10^{-3} {\rm \ Hz}$.\

The typical local magnetic field strength on stellar surfaces ($B_\ast$) is determined by equating gas pressure with magnetic pressure, resulting in an equipartition field strength. Considering a gas composed solely of hydrogen, $B_\ast$ is given by
\begin{align}
    B_\ast = \sqrt{8 \pi \rho_\ast k_{\rm B} T_\ast / m_{\rm H}},
\end{align}
where $m_{\rm H}$ is the mass of a hydrogen atom, assuming nearly zero ionization at the stellar surface. The magnetic field filling factor ($f_\ast$), defined as the ratio of local to average magnetic field strengths, is then given as follows:
\begin{align}
    f_\ast = \langle B_\ast \rangle / B_\ast.
\end{align}

\renewcommand{\arraystretch}{1.25}
\begin{table}[t!]
\caption{Surface parameters adopted in our model.}
\centering
  \begin{tabular}{lll} \hline \hline \\[-1.2em]
    physical quantity & symbol & adopted value \\[0.2em] \hline
    mass density & $\rho_\ast$ & $2.24 \times 10^{-6} {\rm \ g \ cm^{-3}}$ \\
    temperature & $T_\ast$ & $3.80 \times 10^{3} {\rm \ K}$ \\
    magnetic field & $B_\ast$ & $4.21 \times 10^{3} {\rm \ G}$ \\
    magnetic filling factor & $f_\ast$ & 0.713 \\
    correlation length & $\lambda_\ast$ & $1.71 \times 10^{7} {\rm \ cm}$ \\
    convection frequency & $f_\ast^{\rm conv}$ & $2.18 \times 10^{-4} {\rm \ s^{-1}}$ \\
    net Poynting flux & $F_{A,\ast}$ & $1.61 \times 10^9 {\rm \ erg \ cm^{-2} \ s^{-1}}$ \\ \hline \hline
  \end{tabular}
  \label{table:surface_parameters}
\end{table}

The net Poynting flux injected from the photosphere ($F_{A,\ast}$) must be prescribed. In our model, since the Poynting flux is carried by Alfv\'en waves, it is natural to assume that the net energy flux is proportional to the Alfv\'en wave energy flux at the photosphere:\begin{align}
    F_{A,\ast} \propto \rho_\ast \delta v_\ast^2 v_{A,\ast} \propto \rho_\ast^{1/2} \delta v_\ast^2 B_\ast,
\end{align}
where $v_{A,\ast} = B_\ast/\sqrt{4 \pi \rho_\ast}$ is the (local) Alfv\'en speed at the surface.
Utilizing this proportionality allows for an estimation of $F_{A,\ast}$ as follows:
\begin{align}
    F_{A,\ast} &= F_{A,\odot} \left( \frac{\rho_\ast}{\rho_\odot} \right)^{1/2} \left( \frac{\delta v_\ast}{\delta v_\odot} \right)^{2} \left( \frac{B_\ast}{B_\odot} \right) = F_{A,\odot} \left( \frac{\rho_\ast}{\rho_\odot} \right)^{1/3} \left( \frac{T_\ast}{T_\odot} \right)^{19/6},
\end{align}
where we set $F_{A,\odot} = 2.4 \times 10^9 {\rm \ erg \ cm^{-2} \ s^{-1}}$ and $B_\odot = 1.3 \times 10^3 {\rm \ G}$. The derived surface parameters are summarized in Table~\ref{table:surface_parameters}.

\section{Model accuracy across different temperatures \label{app:model_accuracy_vs_temperature}}

\begin{figure}[!t]
\centering
\includegraphics[width=75mm]{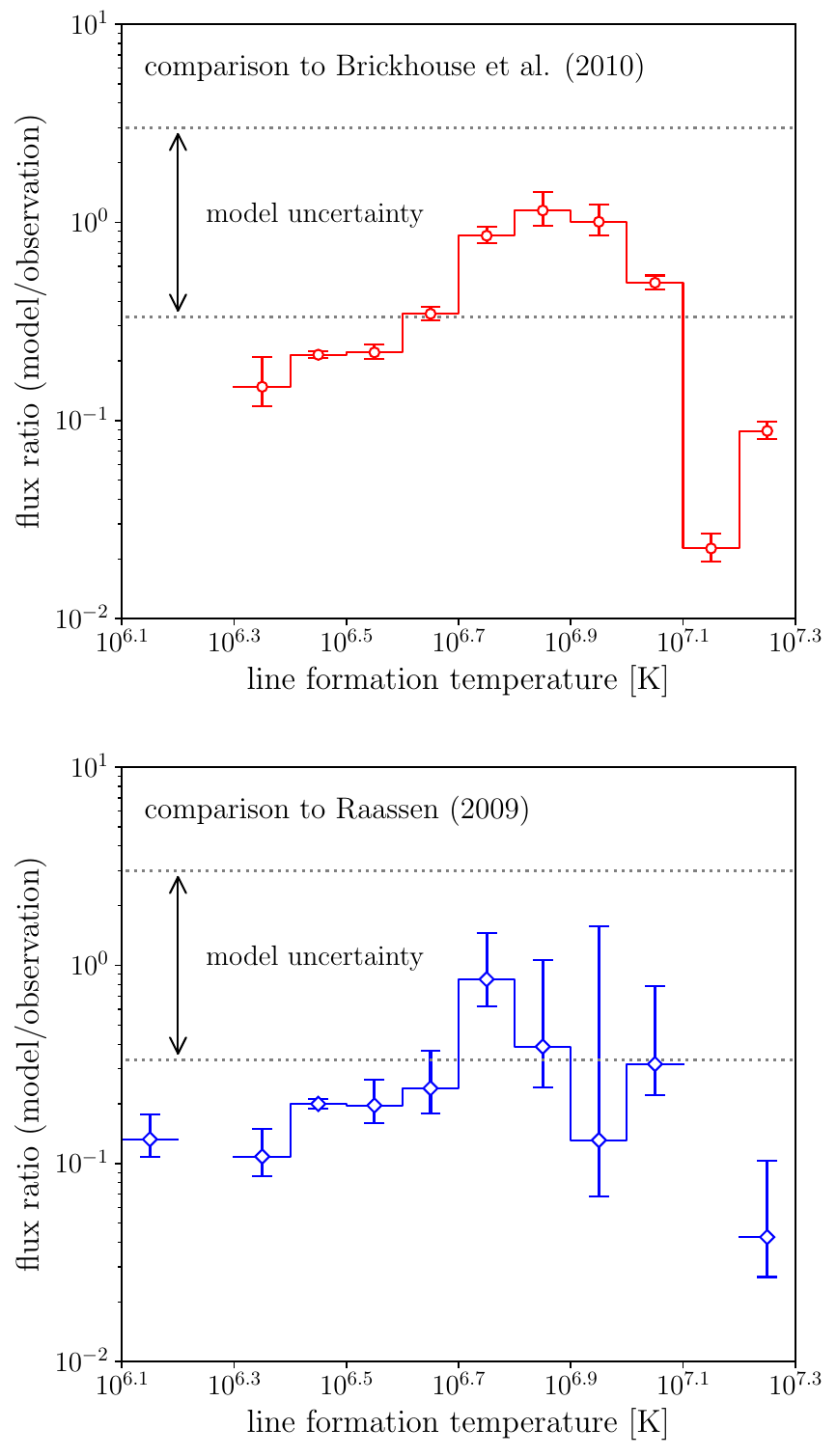}
\caption{
Ratio of modelled photon flux to observed photon flux across various line formation temperatures ($T_{\rm max}$). The top panel compares results with \citet{Brickhouse_2010_ApJ} and the bottom panel with \citet{Raassen_2009_AA}. Ratios are computed for individual emission lines and averaged in temperature bins of 0.1 dex. Error bars indicate 1-sigma observational uncertainties, and dotted lines represent the estimated range of the model uncertainty.
}
\label{fig:formation_temperature_vs_line_accuracy_with_absorption_correction}
\end{figure}

To show the model accuracy in reproducing emission lines at each formation temperature, we show in Fig.~\ref{fig:formation_temperature_vs_line_accuracy_with_absorption_correction} the photon flux ratios between the model and observations. The temperature was binned at 0.1 dex intervals, and the intensity ratios were averaged within each bin. The upper and lower limits of the error bars represent the bin-averaged values calculated by $-1 \sigma$ and $+1 \sigma$ values of the observed data. Since the $-1 \sigma$ value became negative in this analysis, we excluded the data of N VI ${\rm \lambda}$29.535 from \citet{Raassen_2009_AA}.

As suggested by Fig.~\ref{fig:emission_lines_comparison_with_absorption_correction}, the reproducibility of line flux is high for emission lines with formation temperatures near the coronal temperature, particularly in the range of $\log T_{\rm max} = 6.6-7.1$, where the uncertainties in both the observations and the models \citep[0.33–3 times,][]{Shoda_2024_AA} overlap. In contrast, emission lines with $\log T_{\rm max} < 6.6$ can only reproduce $10-30 \%$ of the observed values, and those with $7.1 < \log T_{\rm max}$ can reproduce only $1-10 \%$. These results suggest that accretion shocks and flares are essential for emission in these temperature ranges, while it is also important to note that coronal radiation can contribute a certain percentage (several tens) to emission lines with $\log T_{\rm max} < 6.6$.

\section{Details of the modelled emission lines \label{app:emission_line_list}}

\renewcommand{\arraystretch}{1.10}
\begin{table*}[t!]
\centering
\caption{Observed and modelled X-ray emission lines from TW Hya.}
  \begin{tabular}{cccccc} \hline \hline \\[-1em]
    ion & $\lambda_0$$^{\rm a}$ [${\rm \AA}$] & $\log T_{\rm max}$$^{\rm a}$ & flux$^{\rm b}$ [$10^{-6}$ ph cm$^{-2}$ s$^{-1}$] & flux$^{\rm b}$ [$10^{-6}$ ph cm$^{-2}$ s$^{-1}$] & flux [$10^{-6}$ ph cm$^{-2}$ s$^{-1}$] \\ &  & & \citep{Raassen_2009_AA} & \citep{Brickhouse_2010_ApJ} & (this work) \\[0.0em] \hline
    \\[-0.7em]
    Si XIV & 6.180 & 7.20 & 3.9 $\pm$ 2.3 &  1.87 $\pm$ 0.19 & 0.171 \\ 
    Si XIII & 6.648 & 7.01 & 3.4 $\pm$ 2.2 & 2.72 $\pm$ 0.17 & 1.35 \\
    Si XIII & 6.688 & 6.97 & 2.4 $\pm$ 2.2 & 0.74 $\pm$ 0.14 & 0.326 \\
    Si XIII & 6.740 & 7.00 & 2.7 $\pm$ 2.2 & 1.66 $\pm$ 0.15 & 0.859 \\
    Mg XII & 8.419 & 7.00 & 4.2 & 2.24 $\pm$ 0.20 & 1.22  \\
    Mg XI & 9.169 & 6.82 & 6.4 $\pm$ 3.4 & 2.25 $\pm$ 0.22 & 2.43 \\
    Mg XI & 9.231 & 6.78 &  & 0.90 $\pm$ 0.17 & 0.446 \\
    Mg XI & 9.314 & 6.81 & 12.0 $\pm$ 6.2 & 1.27 $\pm$ 0.19 & 1.65 \\
    Ne X & 9.481 & 6.79 &  & 1.64 $\pm$ 0.20 & 2.28 \\
    Ne X & 9.708 & 6.78 & 7.6 $\pm$ 4.8 & 2.75 $\pm$ 0.27 & 4.02 \\
    Ne X & 10.239 & 6.78 & 11.5 $\pm$ 4.5 & 9.23 $\pm$ 0.53 & 13.0 \\
    Ne IX & 10.764 & 6.62 & 7.7 $\pm$ 3.7 & 3.54 $\pm$ 0.43 & 1.46 \\
    Ne IX & 11.000 & 6.62 & 12.6 $\pm$ 4.2 & 8.48 $\pm$ 0.63 & 3.09 \\
    Ne IX & 11.547 & 6.62 & 22.0 $\pm$ 4.8 & 23.7 $\pm$ 0.99 & 9.58 \\
    Fe XXII & 11.768 & 7.12 &  & 3.28 $\pm$ 0.50 & 0.125 \\
    Fe XXII & 11.934 & 7.12 &  & 3.01 $\pm$ 0.50 & 0.0549 \\
    Ne X & 12.138 & 6.76 & 89.6 $\pm$ 7.8 & 76.1 $\pm$ 1.9 & 108 \\
    Fe XVII & 12.264 & 6.78 &  & 4.3 $\pm$ 0.6 & 3.90 \\
    Ne IX & 13.447 & 6.60 & 215.0 $\pm$ 11.0 & 177. $\pm$ 4.2 & 74.7 \\
    Ne IX & 13.553 & 6.57 & 129.0 $\pm$ 9.0 & 114. $\pm$ 3.3 & 12.4 \\
    Ne IX & 13.699 & 6.59 & 81.2 $\pm$ 7.7 & 58.7 $\pm$ 2.3 & 55.9 \\
    Ni XIX & 14.040 & 6.85 & 5.6 $\pm$ 4.4 &  & 9.46 \\
    Fe XVIII & 14.209 & 6.90 &  & 4.59 $\pm$ 0.76 & 14.0 \\
    Fe XVIII & 14.258 & 6.89 &  & 1.63 $\pm$ 0.53 & 2.78 \\
    O VIII & 14.821 & 6.51 & 6.9 $\pm$ 2.8 & 5.04 $\pm$ 0.80 & 1.22 \\
    Fe XVII & 15.013 & 6.76 & 35.9 $\pm$ 6.5 & 36.5 $\pm$ 1.8 & 41.3 \\
    O VIII & 15.176 & 6.50 & 9.0 $\pm$ 4.1 & 10.4 $\pm$ 1.2 & 2.72 \\
    Fe XVII & 15.262 & 6.76 & 17.4 $\pm$ 5.1 & 17.4 $\pm$ 1.4 & 11.9 \\
    O VIII & 16.006 & 6.50 & 29.7 $\pm$ 5.4 & 30.6 $\pm$ 2.3 & 8.74 \\
    Fe XVIII & 16.072 & 6.87 &  & 5.17 $\pm$ 1.0 & 7.68 \\
    Fe XVII & 16.776 & 6.74 & 20.4 $\pm$ 4.6 & 25.5 $\pm$ 3.0 & 26.0 \\
    Fe XVII & 17.051 & 6.74 & 22.9 $\pm$ 12.6 & 27.9 $\pm$ 2.3 & 33.7 \\
    Fe XVII & 17.096 & 6.72 & 15.1 $\pm$ 10.3 & 26.3 $\pm$ 2.3 & 27.3 \\
    O VII & 17.396 & 6.35 &  & 4.68 $\pm$ 1.5 & 0.513 \\
    O VII & 17.768 & 6.35 &  & 3.66 $\pm$ 1.6 & 1.06 \\
    O VII & 18.627 & 6.34 & 17.5 $\pm$ 6.7 & 17.9 $\pm$ 1.4 & 3.26 \\
    O VIII & 18.967 & 6.48 & 229.0 $\pm$ 12.0 & 213. $\pm$ 8.4 & 71.2 \\
    N VII & 20.910 & 6.33 & 20.7 $\pm$ 8.6 & 3.8 $\pm$ 2.5 & 2.11 \\
    O VII & 21.602 & 6.32 & 119.0 $\pm$ 12.0 & 117. $\pm$ 10.0 & 25.2 \\
    O VII & 21.804 & 6.30 & 103.0 $\pm$ 11.0 & 72.4 $\pm$ 9.1 & 5.16 \\
    O VII & 22.098 & 6.32 & 15.3 $\pm$ 6.3 & 15.2 $\pm$ 4.4 & 20.5 \\
    N VII & 24.779 & 6.31 & 81.2 $\pm$ 11.3 & 68.5 $\pm$ 7.6 & 17.3 \\
    N VI & 28.787 & 6.16 & 31.2 $\pm$ 12.0 &  & 4.72 \\
    N VI & 29.084 & 6.14 & 27.1 $\pm$ 10.0 &  & 1.53 \\
    N VI & 29.535 & 6.15 & 6.0 $\pm$ 6.3 &  & 3.31 \\
    C VI & 33.734 & 6.12 & 59.3 $\pm$ 11.0 &  & 69.81 \\[0.1em]
    \hline \hline
  \end{tabular}
  \tablefoot{$^{\rm a}$ Reference wavelength and the decimal logarithm of temperature (in K) of maximum emissivity from CHIANTI atomic database version 10. $^{\rm b}$ Observed fluxes at Earth with statistical 1$\sigma$ errors.
  }
  \label{table:emission_line_list}
\end{table*}

\begin{figure*}[!t]
\centering
\includegraphics[width=170mm]{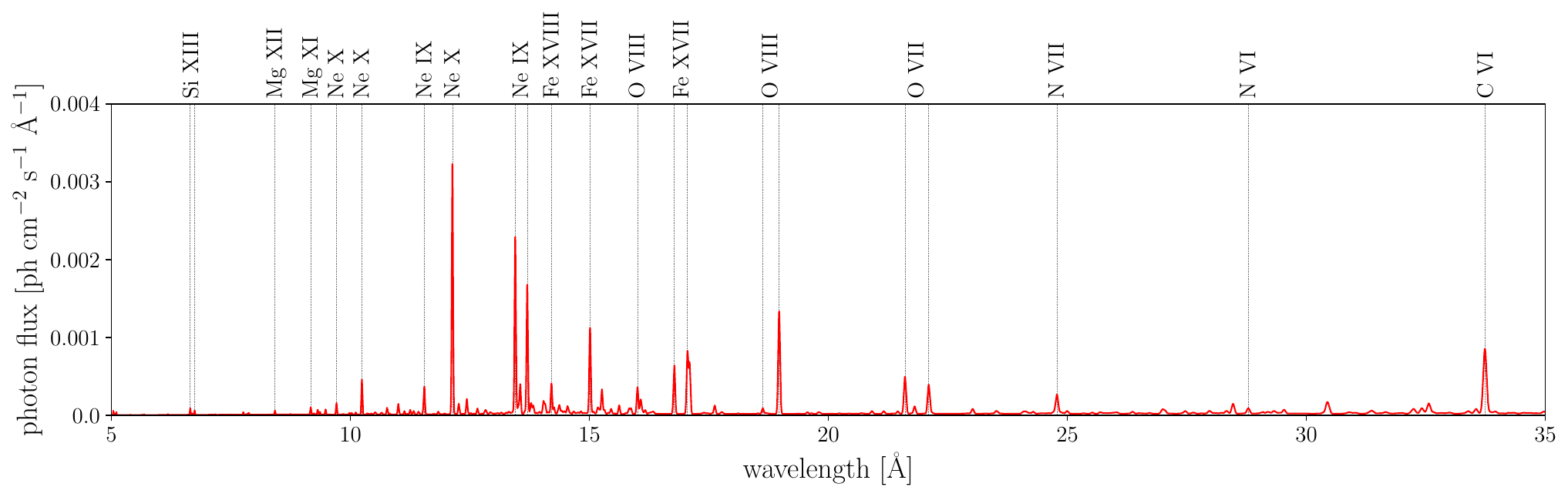}
\caption{
Model spectrum presented as a function of wavelength in the X-ray range. The vertical lines indicate the central wavelength of each emission line, with the corresponding ion labeled above.
}
\label{fig:spectrum_comparison_emission_lines_highlighted}
\end{figure*}

Table~\ref{table:emission_line_list} summarizes the photon fluxes of the modelled emission lines (indicated by the vertical lines in the bottom panel of Fig.~\ref{fig:spectrum_overview}) together with the photon fluxes obtained by \textit{Chandra}/LETG \citep{Raassen_2009_AA} and \textit{Chandra}/HETG \citep{Brickhouse_2010_ApJ}. The corresponding ion species, wavelengths ($\lambda_0$), and temperatures of maximum line emissivity ($\log T_{\rm max}$) are also listed. Emission lines with a significantly low contribution function (Ni XVIII ${\rm \lambda}$14.370, O VII ${\rm \lambda}$17.200) are excluded from the list due to potential misinterpretation of observation or inaccurate calculation of the contribution function. 

Figure~\ref{fig:spectrum_comparison_emission_lines_highlighted} displays the unbinned model spectrum in the X-ray range to provide a more intuitive representation of the emission lines from our model. Each emission line is highlighted by a vertical dashed line with a corresponding ion labeled above. The model values in Table~\ref{table:emission_line_list} represent the photon flux of each emission line in Fig.~\ref{fig:spectrum_comparison_emission_lines_highlighted}

\end{document}